# The bermions: an approach to lattice QCD dynamical fermions from negative flavour numbers

G.M. de Divitiis, R. Frezzotti, M. Guagnelli, M. Masetti
and R. Petronzio

Dipartimento di Fisica, Università di Roma *Tor Vergata*

and

INFN, Sezione di Roma II

Viale della Ricerca Scientifica, 00133 Roma, Italy

July 5, 1995

## Abstract

We estimate the effects of dynamical fermions by extrapolating to positive flavour numbers the results from negative values obtained by adding to the pure gauge sector a fermion action where the fields obey a Bose statistics: the bermions.

# 1 Introduction

Dynamical fermion simulations require the calculation of statistical averages with the pure gauge measure modified by the fermion determinant. Its evaluation is at present the most time consuming step of lattice QCD simulations. In spite of the effort spent to evaluate quantum averages in the full theory, the results from these calculations when compared with those of the quenched approximation, where the fermion determinant is set to unity, appear to amount mainly to a shift in the values of the bare coupling $\beta$. The size of the effect can be estimated by requiring that the lattice spacing of the pure gauge theory ($G$) matches the one of the full theory ($GF$ = Gauge + Fermions) in perturbation theory:

$$a^{GF}(\beta^{GF}) = a^{G}(\beta^{G})$$
$$a^{i}\Lambda = e^{-\frac{\beta}{12 b_0^i}} (6 b_0^i / \beta)^{-\frac{b_1^i}{2(b_0^i)^2}} \qquad\qquad i = GF, G \qquad (1)$$

where $\Lambda$ is the same in the two theories because physical quantities are set to common values and $b_0$ and $b_1$ are the coefficients of one and two loops beta function. What remains of fermion dynamics after the shift appears to be a perturbative correction to the quenched result: in this paper we argue that such a correction can be extrapolated from a theory with negative flavour numbers where the fermion fields are replaced by fields with Bose statistics and fermion action that we call *bermions*. The extrapolation from negative flavour numbers was proposed long ago in ref. [1] for two dimensional models, but the method worked only for heavy quark masses. The results described in this paper have been obtained with a Wilson fermion action corresponding to even flavour numbers. This restriction is due to the necessity of dealing with a positive definite quadratic operator for the bermions which are thermalized together with gauge fields in the Monte Carlo simulation. It is the latter possibility which makes the method very competitive with respect to the best fermion determinant algorithms: we have estimated, for the choice of volume and parameters of this paper, a gain by a factor approximately one hundred with respect to a hybrid Monte Carlo simulation.

In the first part of the paper after a presentation of some technical details we discuss how the extrapolation works when the shift is expected to be perturbative, i.e. for heavy quark masses or in the asymptotic scaling region; in the second part we show how, even for a non perturbative shift, the extrapolation remains smooth when it is performed at constant values of the lattice spacing and of the renormalized quark mass.

## 2 The bermion action

Following Lüscher in ref. [2] we use the action

$$S[U, \phi] = S_g[U] + \sum_x \phi^\dagger(x) Q^2 \phi(x) \qquad (2)$$

where $S_g$ is the standard Wilson action for the gauge part, $Q$ is an operator related to the lattice Dirac operator $D$ by

$$Q = \gamma_5 (D + m)/M, \qquad (3)$$

and $\phi(x)$ is the bermion field.
The constant $M$ is related to $m$ by:

$$M = 8 + m \qquad (4)$$

Introducing the Wilson hopping parameter

$$K = (8 + 2m)^{-1} \qquad (5)$$

and the constant

$$c_0 = (2KM)^{-1} = (1 + 8K)^{-1} \qquad (6)$$

one obtains the result:

$$[Q\phi](x) = c_0 \gamma_5 \{\phi(x) - K \sum_{\mu=0}^{3} [U_\mu(x)(1 - \gamma_\mu)\phi(x + \mu) \\ + U_\mu^\dagger(x - \mu)(1 + \gamma_\mu)\phi(x - \mu)\} \qquad (7)$$

The form of the *bermion* action is suitable for a simulation with a standard heat–bath algorithm. Actually we did use a hybrid algorithm which involves a mixture of heat–bath and of overrelaxation on the gauge fields and on the *bermion* fields. In the SU(3) case we used the Cabibbo-Marinari pseudo–heat–bath algorithm. A full sweep of the lattice is composed of one heat–bath and $N_b$ overrelaxation sweeps over the bermion fields, followed by one pseudo–heat–bath and $N_g$ pseudo–overrelaxation sweeps over the gauge fields. The parameters $N_b$ and $N_g$ are optimized in order to keep auto–correlation times as low as possible. For hadron spectroscopy, we need the inversion of the quark propagator that we perform with the conjugate gradient algorithm every 1000 full sweeps.

## 3 Perturbative dynamical fermions

When the shift is perturbative, the connection between negative and positive flavour numbers can be obtained by a fermion loop expansion [3] and can be expressed in a simple analytic form. If we represent the fermion average as:

$$\langle \mathcal{O} \rangle_{n_f}^F = \frac{\int D[A] e^{-S[A]} \mathcal{O}[A] e^{(n_f/2)\Delta[A]}}{\int D[A] e^{-S[A]} e^{(n_f/2)\Delta[A]}} \quad (8)$$

the corresponding *bermion* average is expressed by:

$$\langle \mathcal{O} \rangle_{n_b}^B = \frac{\int D[A] e^{-S[A]} \mathcal{O}[A] e^{-n_b \Delta[A]}}{\int D[A] e^{-S[A]} e^{-n_b \Delta[A]}} \quad (9)$$

where $n_f$ and $n_b$ are both positive and integer numbers and $\Delta = Tr(\log(Q^2))$. From bermion averages one can produce fermion averages which are correct up to a definite number of fermion loops. For example, an average valid to one loop can be written as:

$$\langle \mathcal{O} \rangle_2^F \overset{1 \text{ loop}}{\sim} \frac{(\langle \mathcal{O} \rangle_0)^2}{\langle \mathcal{O} \rangle_1^B} \quad (10)$$

Similarly, one can build expressions corrected up to a higher loop number:

$$\langle \mathcal{O} \rangle_2^F \overset{2 \text{ loop}}{\sim} \frac{(\langle \mathcal{O} \rangle_0)^3 \langle \mathcal{O} \rangle_2^B}{(\langle \mathcal{O} \rangle_1^B)^3}$$

$$\langle \mathcal{O} \rangle_2^F \overset{3 \text{ loop}}{\sim} \frac{(\langle \mathcal{O} \rangle_0)^4 (\langle \mathcal{O} \rangle_2^B)^4}{(\langle \mathcal{O} \rangle_1^B)^6 \langle \mathcal{O} \rangle_3^B} \quad (11)$$

The equations above are expected to converge rapidly to the unquenched result for heavy quark masses. In table 1 we give the value of the average plaquette for the SU(2) case at a value of $\beta$ far from the asymptotic scaling region at two values of the Wilson parameter $K$. For the first rather low value, i.e. for heavy quark masses, the jump in the plaquette is obtained correctly by using the simple one loop formula in eq.(10). For the second $K$ value, i.e. for lighter quarks, the jump increases and the one loop approximation becomes inadequate. The inclusion of higher loops lowers the discrepancy with the right value but the error of the estimate increases. This loss of precision can be understood by considering that equations (11) imply a ratio of several correlations with a natural increase of the resulting statistical fluctuations.

The steeper flavour number dependence for lighter quarks can be seen from figure 1 where we plot the average plaquette for $\beta = 1.75$ and for different values of the quark mass. The effect can be attributed to the increase in the shift of the $\beta$ values when quarks are lighter and make the contribution of fermion loops fully active; the asymmetry with respect to zero flavour values reflects the different slope of the average plaquette at $\beta$ values lower and higher than the one of the pure gauge case.

A fast convergence is also expected in the asymptotic scaling limit at very large values of $\beta$. We report the perturbative expressions in SU(3) for the average plaquette $\langle P \rangle$ and for the quark mass:

$$1 - \langle P \rangle = \frac{2}{\beta} + \frac{1.2208}{\beta^2} - 0.0588 \frac{n_f}{\beta^2}$$
$$am_q = \ln(1 + \frac{1}{2K} - \frac{1}{2K_c}) = \ln(1 + \frac{1}{2K} - 4 + \frac{5.457 \cdot 3}{2\beta\pi}) \qquad (12)$$

In eq.(12) the fermion mass in plaquette expansion is taken to be zero (see ref. [4]).

From now on our results refer to the SU(3) case only.

We first explore the case of $\beta \simeq 24$ on a $8^3 \cdot 16$ lattice, deep in the asymptotic scaling region. In figure 2 we report the $1 \times 1$ and $2 \times 2$ plaquettes as a function of the flavour number: the result are mildly dependent upon the flavour number. Figure 3 is a blow up of the behaviour of the $2 \times 2$ plaquette: in such an enlarged scale the flavour dependence is visible and appears well fitted by a straight line. The lowest order of the perturbative expansion proportional to the flavour number is adequate at such a high $\beta$ value and provides a nice test of the bermion Monte Carlo.

At lower values of $\beta$ the perturbative expansion breaks down: Parisi has suggested [5] that its validity may be extended by a suitable choice of the expansion parameter eliminating tadpole type corrections which are typical of lattice perturbative calculations and responsible for the large ratio of the lattice $\Lambda$ parameter with respect to, for example, the one of dimensional regularization. The original proposal was to adopt an expansion parameter related to the value of the $1 \times 1$ plaquette. This suggestion was later refined by Lepage and Mackenzie [6] who proposed an improved definition of the coupling constant still related to the $1 \times 1$ plaquette. In the same paper they proposed an expression for the quark mass in terms of the improved coupling evaluated at a different and optimized scale. Comparing different theories at the same value of the improved coupling requires a readjustement of the corresponding $\beta$ values to obtain a matching of the $1 \times 1$ plaquette from which the improved coupling is extracted.

The strategy of matching a physical quantity like the improved coupling at the same lattice spacing scale can be further generalized into the recipe of comparing and then extrapolating theories with different flavour content at fixed values of the renormalized quantities. This procedure reabsorbs the unphysical shift in the bare couplings and allows to extrapolate only the genuine physical effects. In QCD with flavour degeneracy there are two independent quantities which need a renormaliza-tion: the coupling constant and the quark mass. We then need two independent quantities to match in order to keep renormalized quantities fixed. At perturbative level these can be identified with the improved coupling and the quark mass defined in ref. [6]. The effect of the matching is visible also in the perturbative region: in figure 4 we report the variation with the flavour number of the $2 \times 2$ plaquette to be compared with the one in figure 3. The behaviour is still well described by a linear fit, but the variation itself has been reduced by a factor 4.

## 4 Non perturbative dynamical fermions

The procedure described in the last section can be fully appreciated at moderate values of $\beta$: we present the results at fixed bare parameters for $\beta = 5.4$ and $K = .162$ on a $16^3 \cdot 32$ volume in figure 5, where now the flavour dependence is visible also with a poor scale resolution. After the matching described above, the same figure turns into figure 6 where the variations are much smaller and linear. A magnified scale allows to better appreciate the change in the behaviour: we give in figure 7 the $2 \times 2$ plaquette and in figure 8 the $3 \times 3$ plaquette with and without the matching. The values of the bare parameters used are: $\beta = 5.662$ (pure gauge), $\beta = 5.975$ - $K = 0.1676$ ($n_b = 1$), and $\beta = 6.369$ - $K = 0.1722$ ($n_b = 2$).

The interesting question is whether such a procedure can be applied successfully to more physical quantities. We have explored the hadron spectrum on a $16^3 \cdot 32$ lattice. In this case the two independent matching quantities have been chosen to be the pion and rho masses in order to fix the renormalized quantities at a scale appropriate to the problem. The ratio of the two essentially fixes the renormalized quark mass and the value of the rho mass, which shows only a mild dependence upon the quark mass, the renormalized coupling constant. In reality, they represent some complicated definition of the two renormalized parameters. The values of the bare couplings needed to obtain a matching of pion and rho masses do not differ dramatically from those obtained with the improved perturbative matching discussed in the last section: the latter are often the starting point of the matching search. We have also found that the correlations of the operators used to extract the pion and rho masses at large distances are, at small distances, very useful monitors of the matching. These correlations can be obtained in the standard way by inverting quark propagators and assembling them in the correlation and also extracted directly from bermion fields correlations. We have extracted hadron masses using pointlike sources with standard and the smeared quark propagators described in ref. [7].
The results for the hadron spectrum are given in figure 9, 10 and 11 where we plot the proton over rho ratio for values of $R2$, the square of the ratio of pion over rho masses, of $\sim 0.68$, 0.61 and of 0.50 and of the rho mass in lattice units of $\sim 0.705$, 0.70 and of 0.66 respectively. The data points referring to the unquenched case are taken from ref. [8]. For the lowest value of $R2$ the unquenched data with approximately matched values of rho and pion are not available. For the heaviest quark mass we show the two unquenched data with the closest matching. In table 2 we report the values of the bare parameters used for the matching of theories with different flavour numbers and those of the corresponding pion, rho and proton masses. As one can see the matching is not perfect. In general the value of the proton over rho ratio increases when the rho mass or the value of $R2$ increase: by taking into account this effect those points which are not well aligned with the others would do it better with a better matching.

|            | $K = 0.12$ | $K = 0.16$ |
|------------|------------|------------|
| quenched   | 0.4271(1)  | 0.4271(1)  |
| unquenched | 0.433(1)   | 0.4607(5)  |
| 1 loop     | 0.4326(2)  | 0.4510(1)  |
| 2 loops    | 0.4330(4)  | 0.4534(2)  |
| 3 loops    | –          | 0.4542(7)  |

Table 1: Average plaquette for SU(2) with Wilson fermions at $\beta = 1.75$ and $K = 0.12$ (on a $4^4$ lattice), $K = 0.16$ (on a $8^4$ lattice) for the quenched case, the unquenched one with $n_f = 2$ (from HMCA simulations) and the approximated results corrected to a given order in the fermion loop expansion.

Within our statistical errors and the systematic error due to an imperfect matching, there are no sign of physical dynamical fermions effects in the proton over rho ratio down to $R2$ of order 0.5. This conclusion is fully consistent, and actually better supported by the smaller statistical errors of our data, with the published unquenched data. Recent analysis of accurate quenched data [9] show that, by extrapolating to zero lattice spacing and infinite volume, one obtains a value for the proton over rho mass ratio close to the experimental value within at most ten percent, leaving very little room for dynamical fermion effects.

In table 3 we compare the values of $\beta$ used for the matching with those that one would get for different flavour numbers according to the perturbative eq.(1) where the input is the $\beta$ value of the pure gauge case and the output the $\beta$ values of the bermion theories. We also show the values of the non integer flavour number which would bring the perturbative shift into agreement with the one determined by pion and rho matching. In the last two columns of the same table we repeat the same exercise by replacing the bare couplings in eq.(1) with those of the Parisi scheme given by $\beta_P = \beta \langle P \rangle$. In this case and for light quark masses the effective flavour number agrees with the expected integer value. We notice that the effective flavour number corresponding to two bermions ($n_f = -4$) is always roughly twice the one

| $n_f$ | $\beta$ | k | $n_{conf}$ | $m_\pi$ | $m_\rho$ | $M_N$ | $m_\pi^2/m_\rho^2$ | $M_N/m_\rho$ |
|---|---|---|---|---|---|---|---|---|
| -4 | 6.4 | 0.155 | 26 | 0.577(4) | 0.705(6) | 1.12(3) | 0.670(10) | 1.59(4) |
| -2 | 6.1 | 0.1557 | 20 | 0.586(3) | 0.706(7) | 1.120(20) | 0.689(11) | 1.586(25) |
| 0 | 5.767 | 0.1582 | 20 | 0.591(5) | 0.710(7) | 1.125(18) | 0.693(11) | 1.585(24) |
| 2 | 5.5 † | 0.158 | | 0.568(5) | 0.672(9) | 1.095(20) | 0.714 | 1.63(4) |
| 2 | 5.6 † | 0.156 | | 0.457(5) | 0.55(1) | 0.87(2) | 0.690 | 1.58(7) |
| -4 | 6.4 | 0.158 | 36 | 0.549(3) | 0.722(7) | 1.144(15) | 0.574(15) | 1.584(22) |
| -2 | 6.1 | 0.158 | 40 | 0.534(3) | 0.687(8) | 1.079(14) | 0.604(8) | 1.571(20) |
| 0 | 5.7 * | 0.163 | 219 | 0.562(2) | 0.719(4) | 1.138(10) | 0.611 | 1.583(15) |
| 2 | 5.4 † | 0.162 | | 0.570(10) | 0.717(14) | 1.15(2) | 0.631 | 1.60(6) |
| -4 | 6.463 | 0.157 | 45 | 0.478(4) | 0.663(7) | 1.049(20) | 0.520(10) | 1.582(27) |
| -2 | 6.1 | 0.161 | 65 | 0.464(4) | 0.658(8) | 1.037(15) | 0.497(7) | 1.576(20) |
| 0 | 5.7 | 0.165 | 73 | 0.460(4) | 0.658(11) | 1.028(18) | 0.489(10) | 1.562(31) |

Table 2: A summary of hadron masses on a $16^3 \cdot 32$ lattice used in the extrapolation procedure. *Quenched results for $\beta = 5.7$ and $k = 0.163$ are obtained by interpolating the results for $k = 0.160$ and $k = 0.165$ at the same $\beta$ from ref.[8]. †The results for unquenched QCD with 2 dynamical fermions are from ref.[7].

| $n_f$ | $\beta$ | $K$ | $\beta_{pert}$ | $n_f^{eff}$ | $\beta_{pert}^{Par}$ | $n_{f\,Par}^{eff}$ |
|---|---|---|---|---|---|---|
| 0 | 5.7 | .165 | 5.7 | 0 | 5.7 | 0 |
| -2 | 6.1 | .161 | 6.38 | -1.16 | 6.11 | -1.98 |
| -4 | 6.46 | .157 | 7.04 | -2.23 | 6.48 | -3.95 |
| 0 | 5.767 | .1582 | 5.767 | 0 | 5.767 | 0 |
| -2 | 6.1 | .1557 | 6.46 | -.96 | 6.25 | -1.52 |
| -4 | 6.4 | .155 | 7.13 | -1.84 | 6.74 | -2.9 |

Table 3: Comparison between the values of the bare parameters needed for the matching and those estimated from eq.(1)

of the one bermion case supporting the notion of an "effective flavour number".

The introduction of negative flavours may have important consequences on the U(1) problem in QCD [10]: in particular the contribution proportional to the anomaly, splitting the masses of flavour singlet and non singlet pseudoscalars, has an opposite sign in the bermion theory with respect to the the fermion theory, leading to a splitting in the opposite direction, i.e. *lowering* the singlet mass. At very low quark masses this effect may lead to a massless particle while the pion is still massive. We have monitored this phenomenon by calculating flavour singlet and non singlet correlations of bermion fields: we have seen a splitting between the corresponding masses leading to a lower singlet mass. The simulations with almost massless quarks may be problematic due to this effect and may represent an ultimate limitation of the method. At the lowest value of the parameter $R2$ that we have explored ($R2 \simeq 0.35$) and for $n_b = 2$ where the effect is bigger, we have estimated a pion mass of $\simeq 0.4$ and a flavour singlet mass of $\simeq 0.2$ in lattice units, away from critical values. All other simulations were performed under safer conditions, i.e. at heavier quark masses and/or smaller flavour number. We will present elsewhere a detailed analysis of the singlet mass with the bermion method as well of other quantities like the light pseudoscalar and vector decay constants where a visible contribution of sea quarks is expected to bring the quenched lattice data into agreement with experimental values.

The extrapolations from negative to positive flavour numbers is an economical way of estimating dynamical fermion effects. The extrapolation remains smooth when the comparison among theories with different flavour content is performed at fixed renormalized quantities, i.e. after matching two independent physical quantities.

**ACKNOWLEDGEMENTS** The numerical work described in this paper was performed on a 25 Gigaflops machine of the APE100 series. We are grateful to the APE group and in particular F. Marzano, M. Torelli and P. Vicini for their help in installling the machine and solving some minor problems. We also thank the LABEN engineers for their hardware assistance.

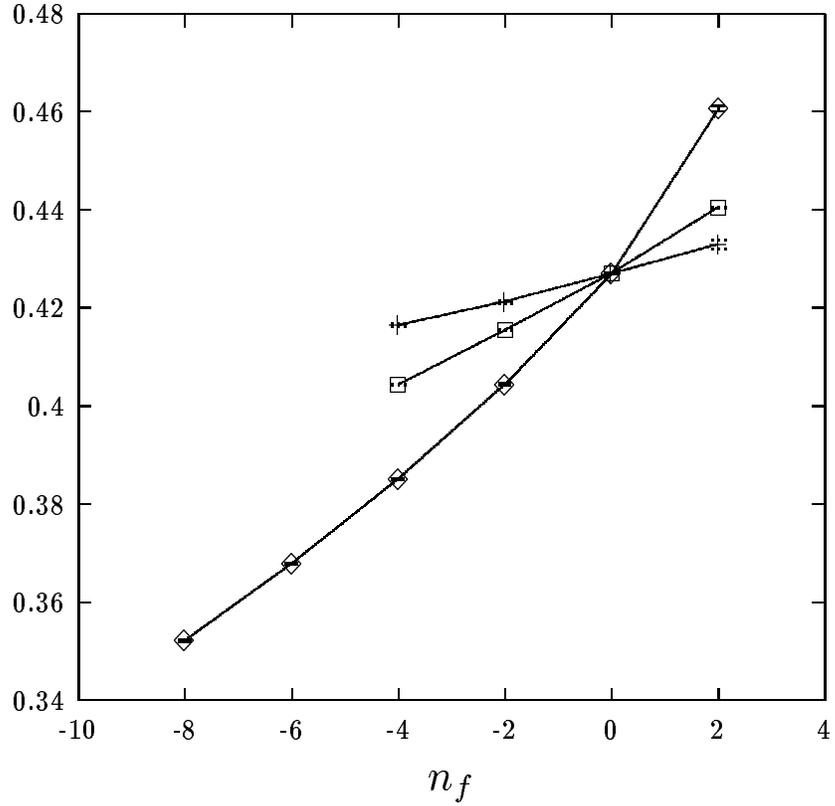

Figure 1: The average plaquette for $SU(2)$ at $\beta = 1.75$ for different values of the Wilson parameter $K = 0.12, 0.14, 0.16$ as a function of the flavour number. The steepest curve refers to the lightest quark mass ($K = 0.16$).

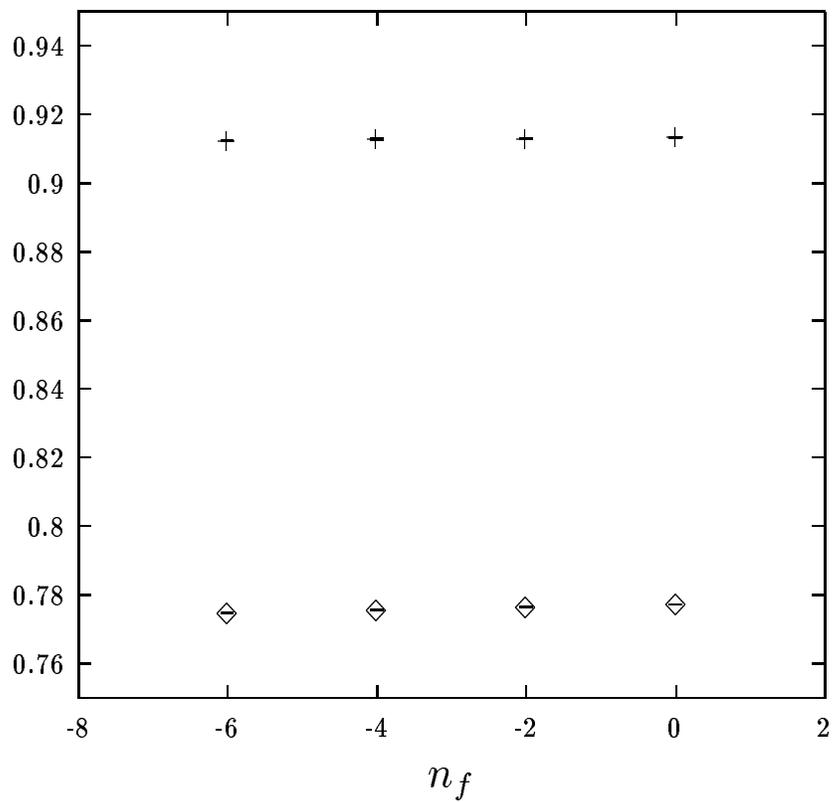

Figure 2: The $1 \times 1$ and $2 \times 2$ average plaquettes for $SU(3)$ at $\beta = 23.76$ and $K = .1285$ corresponding to a negligible ($\sim 10^{-4}$) quark mass as a function of the flavour number.

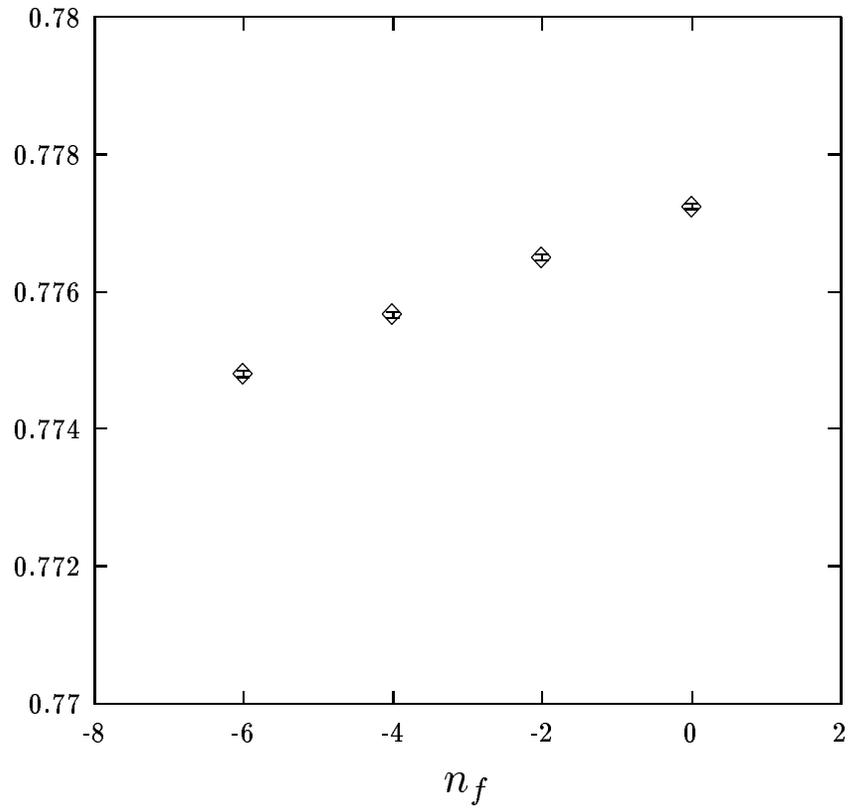

Figure 3: A blow up of the variation of the 2 × 2 average plaquettes as a function of the flavour number, with the same parameters of the previous figure.

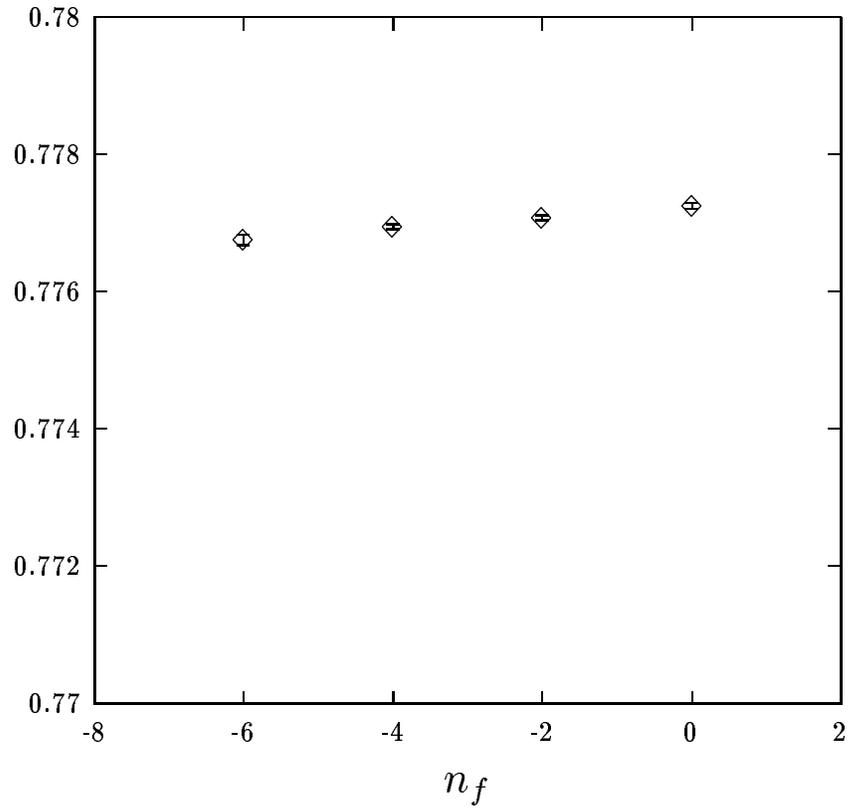

Figure 4: The variation of the 2 × 2 average plaquettes as a function of the flavour number after the matching procedure described in the text. The values of the 1 × 1 plaquette and of the perturbative quark mass of ref. [6] are 0.9134 and less than $10^{-4}$ respectively.

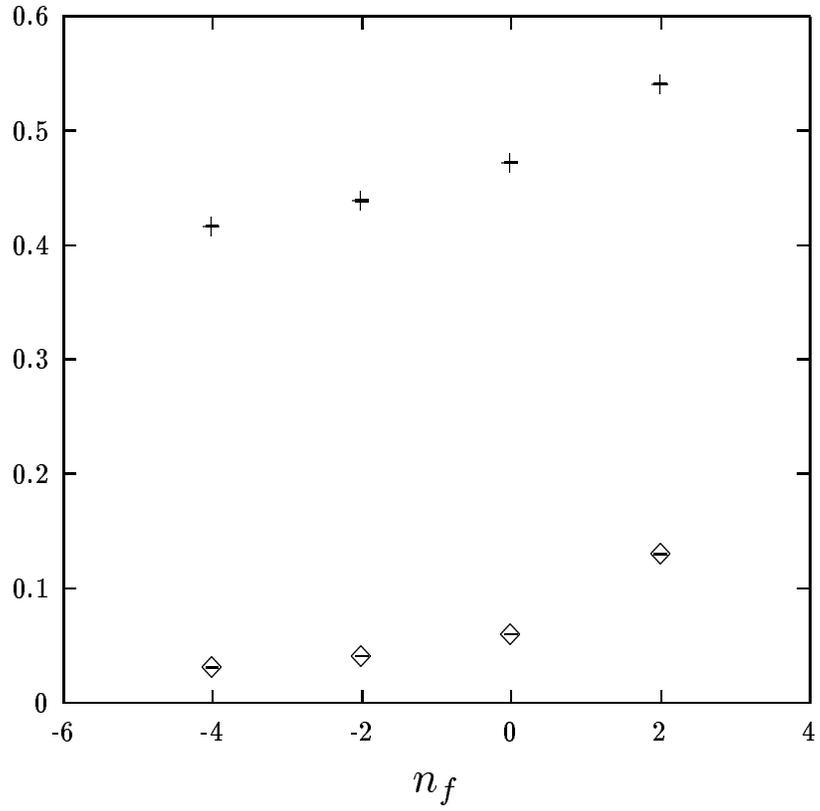

Figure 5: The $1 \times 1$ and $2 \times 2$ average plaquettes for $SU(3)$ at $\beta = 5.4$ and $K = 0.162$ as a function of the flavour number.

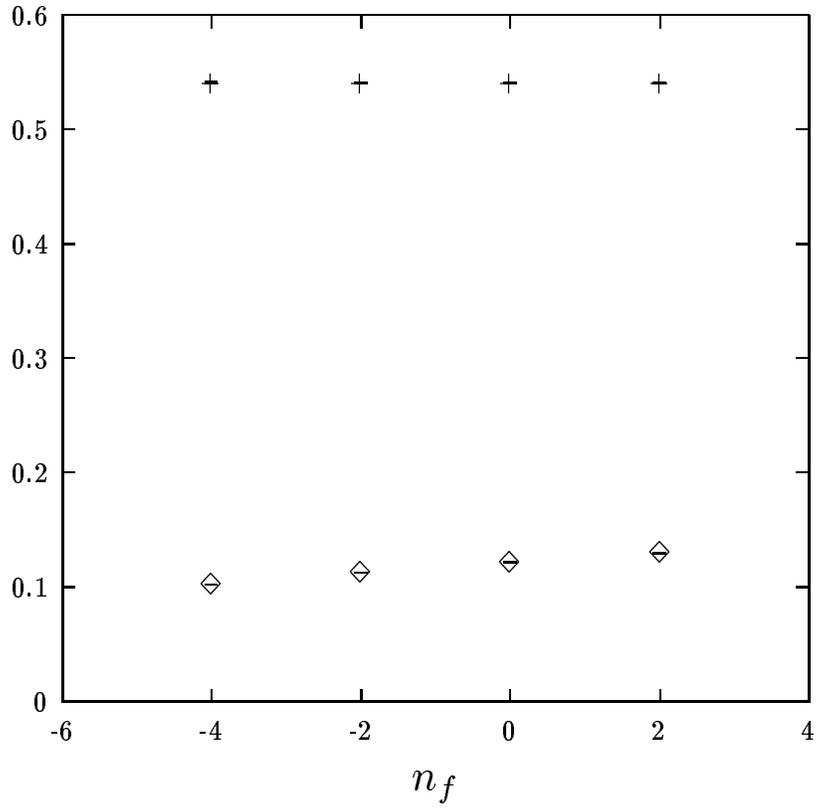

Figure 6: The $1\times 1$ and $2\times 2$ average plaquettes for $SU(3)$ at approximately constant values of the $1 \times 1$ plaquettes and quark mass defined as in ref.[6] of about 0.05 as a function of the flavour number, after the matching procedure described in the text.

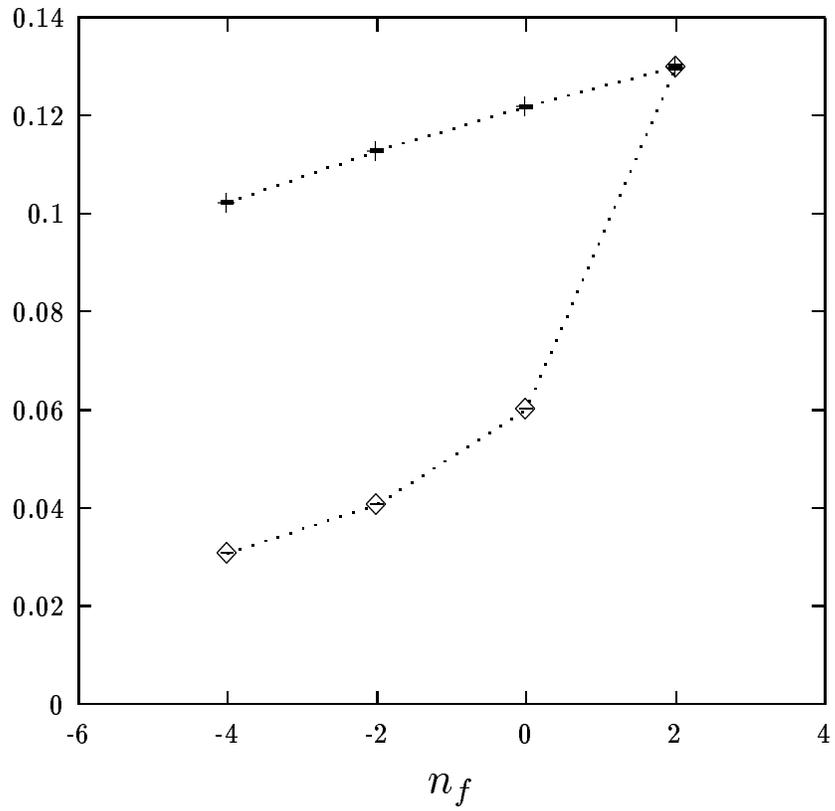

Figure 7: A comparison between the variation with the flavour number of the $2 \times 2$ plaquettes with bare parameters fixed (diamonds) and with the matching (crosses) discussed in the text.

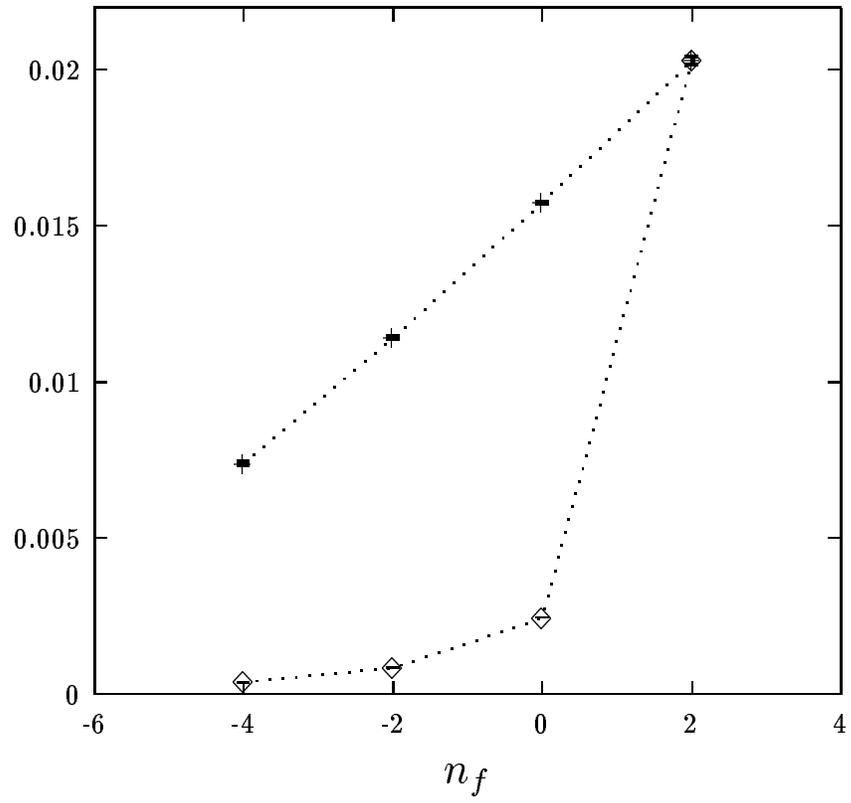

Figure 8: A comparison between the variation with the flavour number of the $3 \times 3$ plaquettes with bare parameters fixed (diamonds) and with the matching (crosses) discussed in the text.

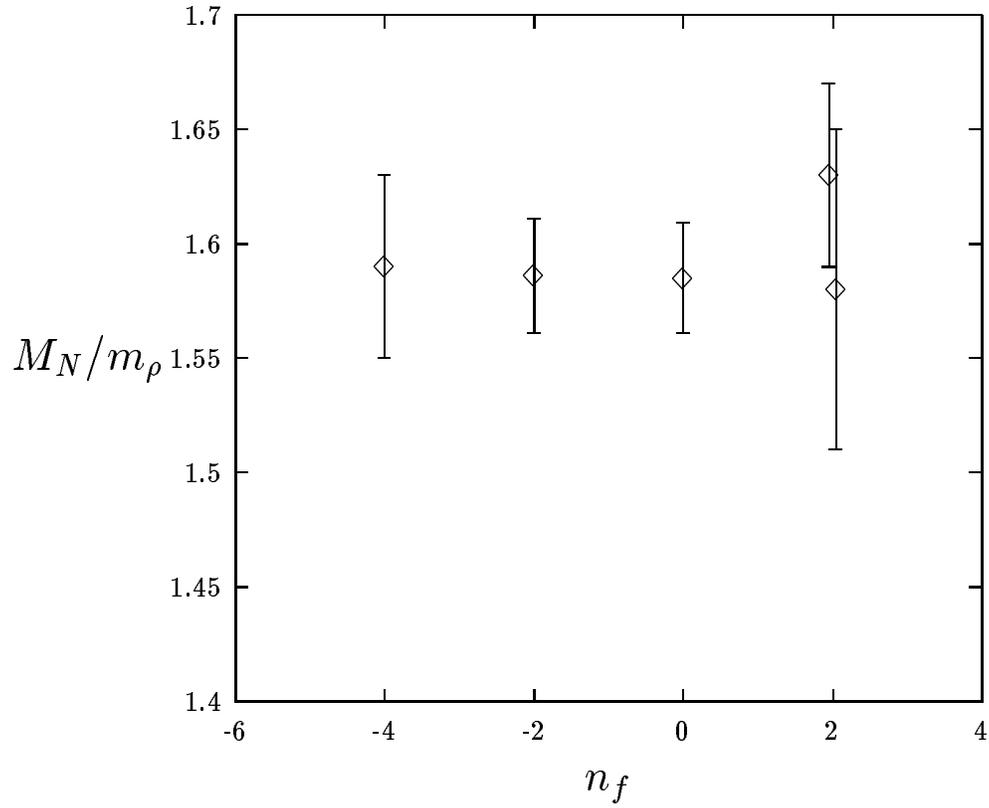

Figure 9: The ratio nucleon over the $\rho$ mass as a function of the flavour number at approximately constant values of the $\rho$ and $\pi$ masses of 0.705 and 0.58 respectively. The corresponding numerical values are given in table 1.

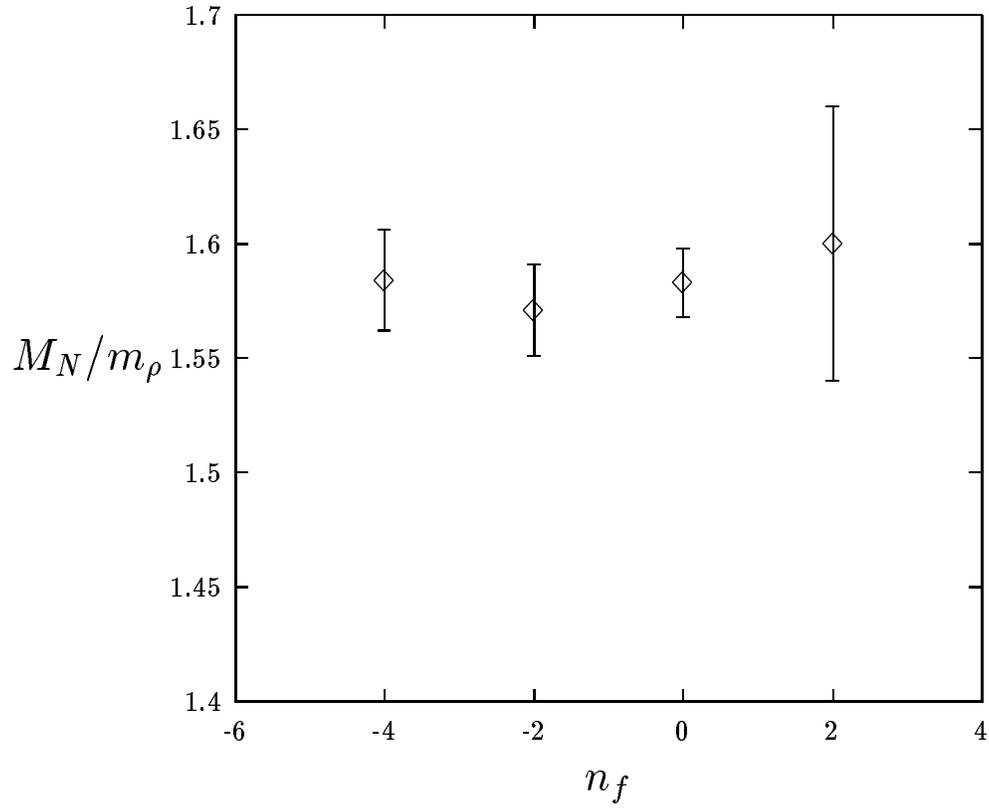

Figure 10: The ratio nucleon over the $\rho$ mass as a function of the flavour number at approximately constant values of the $\rho$ and $\pi$ masses of 0.700 and 0.55 respectively. The corresponding numerical values are given in table 1.

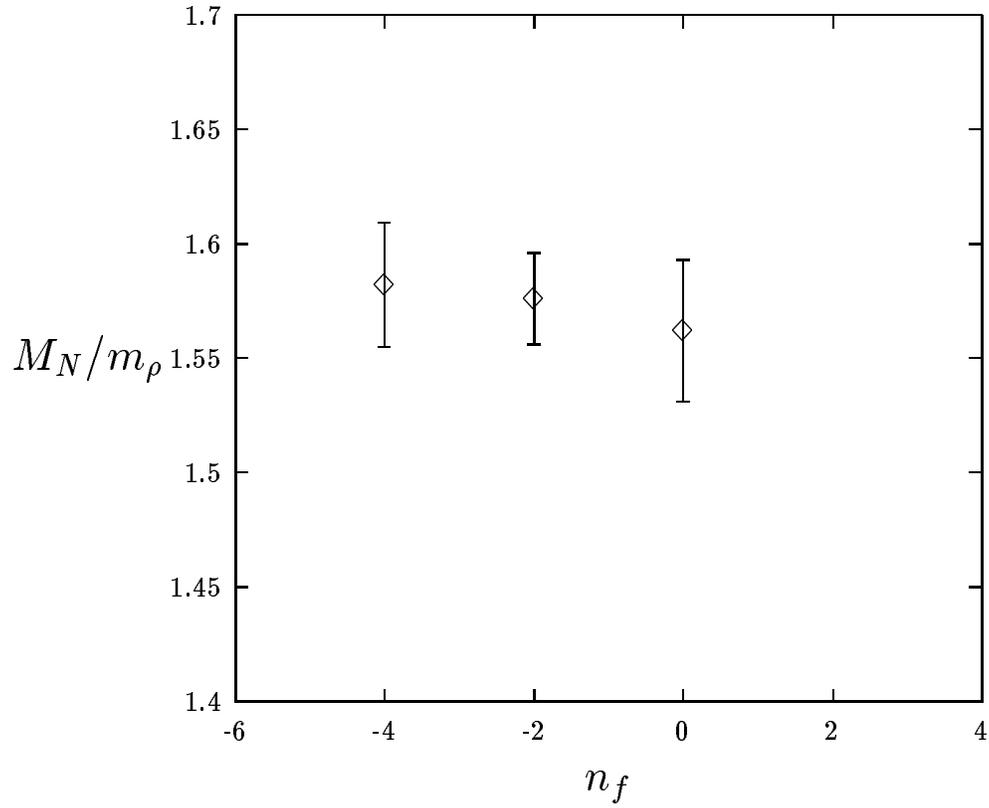

Figure 11: The ratio nucleon over the $\rho$ mass as a function of the flavour number at approximately constant values of the $\rho$ and $\pi$ masses of 0.66 and 0.47 respectively. The corresponding numerical values are given in table 1.